\documentclass[12pt,a4paper]{JHEP3}
\newcommand{\be}{\begin{equation}}
\newcommand{\ee}{\end{equation}}
\newcommand{\bea}{\begin{eqnarray}}
\newcommand{\eea}{\end{eqnarray}}
\newcommand{\non}{\nonumber}

\title{Self-T-Dual Brane Cosmology and the Cosmological Constant
  Problem}

\author{Olindo Corradini
\\Dipartimento  di Fisica, Universit{\`a} di Bologna
\\and  INFN, Sezione di Bologna\\
Via Irnerio, 46 - Bologna I-40126, Italy
\\ E-mail: \email{olindo.corradini@bo.infn.it}}

\author{Massimiliano Rinaldi\\
School of Mathematical Sciences,\\
University College Dublin,\\
Belfield, Dublin 4, Ireland\\
E-mail: \email{massimiliano.rinaldi@ucd.ie}}

\abstract{We consider a codimension-one brane embedded in a
gravity-dilaton bulk action, whose symmetries are compatible with
T-duality along the space-like directions parallel to the brane,
and the bulk time-like direction. The equations of motions in the
string frame allow for a smooth background obtained by the union
of two symmetric patches of AdS space. The Poincar\'{e} invariance
of the solution appears to hold independently of the value of the
brane vacuum energy, through a self-tuning property of the dilaton
ground state. Moreover, the effective cosmology displays a bounce,
at which the scale factor does not shrink to zero. Finally, by
exploiting the T-duality symmetry, we show how to construct an
ever-expanding Universe, along the lines of the Pre-Big Bang
scenario.}

\preprint{} \keywords{String Theory and Cosmology}

\begin{document}

\section{Introduction}\setcounter{equation}{0}

The major problem associated with the Randall-Sundrum (RS) models
\cite{RS1,RS2} is probably represented by the fine-tuning between
the vacuum expectation value of the brane fields and the bulk
cosmological constant, which is required to ensure that the
background is the union of smooth AdS patches.

An attempt to tackle the fine-tuning problem was realised by
Kachru, Schulz, and Silverstein  (KSS) by introducing a bulk
scalar field \cite{KSS} (see also \cite{chamb}). The addition of
this extra degree of freedom lifts the fine-tuning to a less
restrictive self-tuning mechanism, in which the dilaton field
adjusts spontaneously in such a way that Poincar\'{e}
4-dimensional solutions exist, no matter what the value of the
brane vacuum energy is.~\footnote{See~\cite{Krause:2000uj} for a
(incomplete)
  list of interesting
  alternative attempts to solve the fine-tuning problem within the brane-world
  scenario.}

The KSS model suffers from a serious drawback, namely the
existence of naked singularities in the bulk. These, however, can
be treated in a few ways. For example, Low and Zee~\cite{Low} have
shown that upon adding a bulk Gauss-Bonnet combination and a bulk
(super-)potential one can make the singularity disappear, but this
leads again to fine-tuned parameters.~\footnote{A detailed
discussion upon self-tuning models can be found
in~\cite{Csaki:2000wz}.} Alternatively, the naked singularity can
become harmless if one considers time-dependent bulks, as first
noted in \cite{Horowitz} (see also \cite{Langlois}). Indeed, in
this case the singularity becomes null and cannot be reached by
the brane. However, the bulk space-time must be cut off by hand at
the singularity, in order to have a finite effective Planck
constant. The possibility of screening
  the bulk singularity behind a horizon was shown to lead to a no-go theorem,
  unless some space-curvature is given to the brane~\cite{Cline:2001yt}. Also,
  the presence of a
curvature singularity compromises the truncation of the effective
action to its lowest order, and needs the inclusion of bulk
higher-curvature terms.

In this Letter, we reconsider some of these issues from the point
of view of the string frame. The motivation relies on the
assumption that the brane lives in a background obtained by some
compactification of string theory. Therefore, there might be
inherited string dualities, which can play an important r\^{o}le
in the brane dynamics. If these dualities are taken into account,
they should then be manifest in the string frame. In particular,
we consider T-duality as the underlying symmetry that must be
preserved not only by the equation of motions for the bulk, but
also by the junction conditions associated to the brane. This
approach was already adopted in \cite{Max1}, and further developed
in \cite{Max2}, where a strong connection with the Pre-Big Bang
(PBB) scenario was discovered. Here, we extend the results of
\cite{Max2} to the case when T-duality acts also along the bulk
time direction. As explained in more detail below, the time-like
T-duality is motivated by the choice of a warped metric as
background geometry, which is Poincar\'{e} invariant along the
brane coordinates.

It is known that T-duality is a mapping between type IIA and type
IIB string theory. At low energy, this symmetry still holds at the
level of the equations of motion~\cite{Buscher}. In particular,
when all the forms of the RR sector of the theory are switched
off, the action becomes self-T-dual. With this we mean that
T-duality becomes a mapping between different solutions to the
equations of motion derived from the {\it same} action, a fact
which lies at the foundations of the PBB scenario
\cite{GaspRev,Lidsey}.

T-duality along the time direction is known to map type II (A or
B) to type II$^*$ (B or A) string theory \cite{hull}. Also in this
case, when all forms are switched off, type II and type II$^*$
actions and equations of motion become self-dual at the tree
level. In general, time-like T-duality requires the time direction
to be compact, and this can be obtained by means of analytic
continuation (see for example \cite{welch}). In our case, we show
instead that the duality symmetry is meaningful because one of the
possible bulk geometry is actually AdS, which is known to contain
closed time-like curves. This solution is particularly important
because, as we will see, it is Poincar\'{e} invariant along the
brane directions, and it does not require any fine-tuning between
bulk and brane parameters.

The insertion of a codimension-one  brane in the bulk geometry
requires junction conditions, which must be satisfied by the
metric components and by the dilaton, upon specification of the
energy-momentum tensor of the localised matter. These conditions
lead to the effective cosmological equations, which govern the
evolution of the Universe as seen by an observer living on the
brane. In this work, we assume that also these equations must be
invariant under T-duality, and this leads to certain
transformation laws for the brane matter fields. These are
different from the ones found in \cite{Max2} (where only
space-like duality was considered), which are similar to the usual
transformation laws of the PBB scenario.

Throughout this Letter, we refer to the brane embedded in a
self-T-dual bulk as ``self-T-dual brane''. This terminology is
justified by the fact that if we take an NS5-brane as a prototype
for our brane, then T-duality along directions longitudinal to the
world-volume of the brane leaves it unchanged. Indeed, the T-dual
partner is again an NS5-brane wrapped on the dual directions (see
e.g.~\cite{Argurio:1998cp} for a general discussion upon solitonic
objects in string theory).

At the level of the effective brane cosmology, we find both static
and evolving models of the Universe, together with their dual
descriptions. In particular, the dynamical models display the same
bouncing behaviour as their dual counterpart. An alternative to
the bouncing cosmology can be built by ``gluing'' part of the
solution to its dual, along the lines of the PBB scenario. In this
way, one obtains an ever expanding cosmological model. This
construction seems also to suggest the possibility of a
phase-transition between a (pre-bounce) solution with
non-localised brane gravity, and a (post-bounce) solution with
localised gravity. However, there are some caveats about this
construction, which will be pointed out.

The structure of this work is the following: in
Sec.~\ref{gensolutions} we consider the bulk geometry in string
frame, and we show that we can find a self-T-dual AdS background.
In Sec.~\ref{Israel} we study the junction conditions induced by
the insertion of a brane in the bulk. By imposing the
self-T-duality on the junction equations, we find the
transformation laws of the energy density and of the pressure, in
the case when the brane matter is taken to be a perfect fluid.
Equipped with these information, we then turn our attention to the
brane cosmological equations in Sec.~\ref{cosmology}, where we
find the various cosmological solutions mentioned above. We
finally end this Letter with some remarks and open problems.

\section{Background geometry and time-like T-duality}\setcounter{equation}{0}\label{gensolutions}

\noindent We consider the D-dimensional action in string the frame
given by \bea\label{Daction} S=S_{\rm bulk}+S_{\rm
brane}&=&{1\over 2}\int_{\cal
M}d^Dx\sqrt{-g}\,e^{-2\phi}\left[R+4(\nabla
\phi)^2-V\right]+\\\non\\&{}&-\int_{\Sigma}d^{D-1}x\sqrt{-h}\,
e^{-2\phi}\left[K^{\pm}-{\cal L}\right]~,\eea where
$K^{\pm}=K^++K^-$ is the extrinsic curvature on the two sides of
the brane submanifold $\Sigma$ embedded in the bulk spacetime
${\cal M}$. Also, $h_{\mu\nu}=g_{\mu\nu}-n_{\mu}n_{\nu}$ is the
induced metric on $\Sigma$, where $n^A$ are the components of the
normal vector pointing into the bulk. We assume that the bulk
metric has the form \bea\label{metric2}
ds^2=g_{MN}dx^Mdx^N=e^{2\sigma(z)}\eta_{\mu\nu}dx^{\mu}dx^{\nu}+dz^2~,\eea
where $\eta_{\mu\nu}dx^{\mu}dx^{\nu}$ stands for the Minkowski
space-time of dimension $D-1$. This metric is clearly invariant
under translations along $x^{\mu}$. Therefore, by assuming that
the T-duality transformation is allowed along the time coordinate,
we expect the equations of motion (and, eventually, the action) to
be invariant under \bea \sigma(z)\stackrel{\rm T}{\longrightarrow}
-\sigma(z)~,\qquad \phi\stackrel{\rm T}{\longrightarrow}
\phi-(D-1)\sigma~,\eea according to Buscher's rules
\cite{Buscher}. To prove this, we consider first the Ricci scalar,
which reads \bea
g^{MN}R_{MN}=-(D-1)(2\sigma''+D\sigma'^{\,2})~,\eea where the
prime denotes differentiation with respect to $z$. By defining the
shifted dilaton as \bea\label{shifteddil} \bar\phi=\phi-{D-1\over
2}\sigma~,\eea and after partial integration, we can write the
bulk part of the action (\ref{Daction}) as \bea S_{\rm
bulk}&=&{1\over 2}\int_{\cal
M}d^Dx\,e^{-2\bar\phi}\left[4\bar\phi'^{\,2}-(D-1)\sigma'^{\,2}-V\right]+
\\\non\\
&{}&-(D-1)\int_{\partial{\cal M}}\hskip-.25cmd^{D-2}x
dt\,\sigma'e^{-2\bar\phi}~.\eea
If the induced metric has the form \bea
h_{\mu\nu}dx^{\mu}dx^{\nu}=-d\tau^2+e^{2\sigma(z)}\delta_{ij}\,dx^idx^j~,\eea
it follows that $\sqrt{-h}e^{-2\phi}=e^{-\sigma} e^{-2\bar\phi}$,
and the boundary part of the action reads \bea S_{\rm
boundary}=-\int_{\Sigma}d^{D-2}xd\tau\,e^{-\sigma}e^{-2\bar\phi}\,
2K~,\eea where we also assumed a $Z_2$ symmetry along the
transverse direction. With the help of the normalisation condition
\bea dt = e^{-\sigma}\sqrt{1+\dot z^2}d\tau~, \eea we find that
the full action reads \bea S&=&{1\over 2}\int_{\cal
M}d^Dx\,e^{-2\bar\phi}\left[4\bar\phi'^{\,2}-(D-1)\sigma'^{\,2}-V\right]+
\non\\
&&-\int_{\partial{\cal M}}\hskip-.25cm d^{D-2}x dt\,
e^{-2\bar\phi}\left[2(D-1)\sigma'+{2K-{\cal L}\over \sqrt{1+\dot
z^2}}\right]~. \eea Given that, by definition, the shifted dilaton
is unchanged by T-duality, it is evident that the bulk part of the
action is invariant when $V$ is a function of $\bar\phi$ only. The
boundary part is also invariant, as this follows from the
transformation law of the trace of the extrinsic curvature (see
the Appendix)\bea K\stackrel{\rm T}{\longrightarrow}\tilde
K=K+2(D-1)\sigma'\sqrt{1+\dot z^2}~,\eea and from the fact that
$\sqrt{1+\dot z^2}=-n_z$ does not change under T-duality. However,
what really matters here is whether the bulk equations of motion
are invariant under T-duality. This is indeed the case, as it can
clearly be seen from the expressions \bea\label{sigmaeq}
0&=&\bar\phi''-\bar\phi'^{\,2}-{D-1\over 4}\sigma'^{\,2}+{1\over
8}\left({\partial V\over \partial
\bar\phi}-2V\right)~,\\\non\\\label{phieq} 0&=&{d\over
dz}\left(\sigma'\,e^{-2\bar\phi}\right)~,\eea obtained by
variation of the action with respect to $\sigma$ and $\bar\phi$.

These equations can be easily solved if we assume that $V$ is an
exponential function of $\bar\phi$. Potentials of this type were
first introduced in the context of string cosmology in order to
obtain a regular curvature bounce
\cite{Meissner1}-\cite{Gasp1992em}. In general, such potentials
are not invariant under general coordinate transformations.
However, a  covariant formulation for the shifted dilaton (and
hence for the potential) was recently proposed in \cite{Gasp2}. In
these works, it is also argued that exponential potentials of the
shifted dilaton can be generated by string loop corrections, as
$e^{\,2\bar\phi}$ is interpreted as the ``reduced'' - i.e.
one-dimensional - string coupling constant. The generally
covariant shifted dilaton is then defined as \bea
e^{-2\bar\phi}=2\int
d^Dy\sqrt{-g(y)}e^{-2\phi(y)}\sqrt{\partial_{\mu}\phi(y)
\partial^{\mu}\phi(y)}\delta(\phi(x)-\phi(y))~,\eea and the
equations of motion are modified accordingly. In our case, it
turns that the above definition for the shifted dilaton coincides
with Eq.~(\ref{shifteddil}), and the equations of motion are given
by \bea\label{eq:bulk1-1} 0&=&
\bar\phi''-\bar\phi'^{\,2}-{D-1\over
4}\,\sigma'^{\,2}+V_0\,e^{\,\beta\bar\phi}~,\\\non\\
\sigma^{\,\prime}&=&ke^{2\bar\phi},\eea where we assumed that \bea
V={8V_0\over \beta-2}\,e^{\,\beta\bar\phi}~.\eea In these
expressions, $V_0$, $\beta$, and $k$ are constants. We consider
the ansatz solution $\bar\phi=\alpha \ln (z-z_0)+\bar\phi_0$. By
substituting this expression, together with the second equation of
motion, into the first, we find \bea
4\alpha(\alpha+1)+k^2(D-1)(z-z_0)^{\,4\alpha+2}\,e^{\,4\bar\phi_0}-
4V_0(z-z_0)^{\,\alpha\beta+2}\,e^{\,\beta\bar\phi_0}=0~. \eea The
analysis of this equation is straightforward and leads to three
type of solutions, namely \bea\label{bulksolutions}
\begin{array}{lcl}\alpha=-\frac{1}{2}~,&\qquad \beta=4~,&\qquad
k^2={1\over D-1}\left(4V_0+ e^{-4\bar\phi_0}\right)~~,\\\\
\alpha=-1~,&\qquad \beta=4~,&\qquad k^2={4\over D-1}V_0~~,\\\\
\alpha=0~,&\qquad \forall\beta~,&\qquad k^2={4\over
D-1}V_0\,e^{\,(\beta-4)\,\bar\phi_0}~,\end{array}\eea where in the
last two cases $V_0>0~$. By studying the Ricci scalar and higher
order curvature invariants, it turns out that the first two
solutions carry a naked singularity at $z=0$.

The last solution however is singularity-free. Indeed, the
background is AdS as the Ricci scalar reads\bea {\cal R} =
-D(D-1)\lambda^2 <0~, \eea where \bea \lambda= 2 \left({V_0\over
D-1}\right)^{1\over 2}e^{\,\beta\bar\phi_0\over 2}~~.\eea Also,
the shifted dilaton becomes constant, while the warp factor and
the
dilaton read respectively\bea \sigma(z)&=& \pm\lambda (z-z_0)~~,\\\non\\
\phi(z)&=&\pm {D-1\over 2}\lambda (z-z_0)+\bar\phi_0~~,\eea for
arbitrary $z_0$. It is interesting to note that the bulk potential
is now constant, but the AdS solution is insensitive to its sign
(which is determined by the arbitrary quantity $\beta-2$).
However, the value of $\beta$ can be fixed if one allows for
variations of the metric element $g_{zz}$. Then, there is a
further equation of motion, known as the ``zero energy condition"
(see for example \cite{Meissner1}), which reads \bea
4\bar\phi'^{\,2}-(D-1)\sigma'^{\,2}+V=0~.\eea In our case, the net
effect of this constraint is that $\beta=4$, which implies a
positive potential. Notice that $\beta=4$ corresponds to a
two-loop potential in the language of~\cite{Gasp2}.

In summary, the metric and the dilaton field, corresponding to the
solution $\sigma=\lambda(z-z_0)$, are given respectively by \bea
ds^2=e^{2\lambda (z-z_0)}\eta_{\mu\nu}
dx^{\mu}dx^{\nu}+dz^2,\qquad \phi(z)={D-1\over 2}\lambda
(z-z_0)+\bar\phi_0~,\eea  while their dual counterparts,
corresponding to the solution $\tilde\sigma=-\lambda(z-z_0)$, are
\bea d\tilde s^2=e^{-2\lambda (z-z_0)}\eta_{\mu\nu}
dx^{\mu}dx^{\nu}+dz^2,\qquad \tilde\phi(z)=-{D-1\over 2}\lambda
(z-z_0)+\bar\phi_0~.\eea From these expressions, it becomes clear
that the dual counterparts represent the same physical solution,
because the T-duality transformation is in fact equivalent to the
mere change of coordinate $z\rightarrow -z$. However, as we will
see below, this equivalence does not hold when a brane with $Z_2$
symmetry is inserted in the background.

\section{Invariant junction
conditions}\setcounter{equation}{0}\label{Israel}

We now turn our attention to the case when a brane is embedded in
the background. As mentioned in the introduction, the idea is to
impose the invariance under T-duality of the equations of motion
and of the junction conditions. These can be written in string
frame as (see~\cite{chamb,Max1,Barcelo:2000js}) \bea\label{sfcurv}
K_{\mu\nu}&=&-{1\over 2}T_{\mu\nu}-{1\over
4}h_{\mu\nu}\,T^{\phi}~,\\\non\\\label{sfdil}
n^A\partial_A\phi&=&{K\over 2}+{T^{\phi}\over 8}~,\eea where
\bea\label{Tphi} T^{\phi}&=&e^{2\phi}{\delta(e^{-2\phi}{\cal
L})\over \delta\phi}~,\\\non\\\label{Tmunu}
T_{\mu\nu}&=&-{2\over\sqrt{-h}}{\delta\sqrt{-h}{\cal L}\over\delta
h^{\mu\nu}}~.\eea The invariance under T-duality imposes
constraints on the transformation law of $T^{\phi}$ and
$T_{\mu\nu}$. To see this, we write Eq.~(\ref{sfdil}) in terms of
the shifted dilaton (\ref{shifteddil}), and we use the explicit
expression for the trace of $K$ (see Appendix
\ref{appendix:extrinsic}). Thus, we find \bea\label{sfdil2}
n^z\bar\phi'-{1\over 2}{dn^z\over dz}={T^{\phi}\over 8}~.\eea
Since $\bar\phi$ and $n^z$ are invariant under T-duality, we have
that\bea\label{invtfi} T^{\phi}\stackrel{\rm
T}{\longrightarrow}\tilde T^{\phi}=T^{\phi}~.\eea If we assume
that the matter on the brane behaves as a perfect fluid, namely
$T^{\mu}_{\,\,\nu}={\rm diag}(-\mu,p,p,\ldots,p)$ and $\tilde
T^{\mu}_{\,\,\nu}={\rm diag}(-\tilde\mu,\tilde p,\tilde
p,\ldots,\tilde p)$, it follows, from the $ij$ and $\tau\tau$
components of Eq.~(\ref{sfcurv}), that \bea \mu &\stackrel{\rm
T}{\longrightarrow}&\tilde\mu=\mu+2p+T^{\phi}~,\\\non\\
p&\stackrel{\rm T}{\longrightarrow}&\tilde p=-p-T^{\phi}~.\eea
Note that the last two conditions also imply that
\bea\label{ident} \tilde\mu+\tilde p=\mu+p~.\eea Therefore, if
$p=\omega\mu$ and $\tilde p=\tilde\omega\tilde\mu$, and
$\omega=-1$, then $\tilde\omega=-1$ as well, independently of the
value of $T^{\phi}$.

The junction conditions can be written in the form of cosmological
equations by using the explicit components of the extrinsic
curvature, and by defining the Hubble parameter as $H=\dot
z\sigma'=\partial_{\tau}\ln A$, where $A(\tau)$ is the scale
factor. Indeed, from the $ij$ component of Eq.~(\ref{sfcurv}), we
find the effective Friedmann equation \bea\label{longFr}
H^2={1\over 4}\left(p+{1\over 2}T^{\phi}\right)^2-\left({A'\over
A}\right)^2~,\eea while the $\tau\tau$ component of
Eq.~(\ref{sfcurv}) yields the evolution equation of $\mu$ and
$p$\bea (\mu+p)H-(p+{1\over 2}T^{\phi}){d\over d\tau}\ln {A'\over
A}+ {d\over d\tau}\left(p+{1\over 2} T^{\phi}\right)=0~.
\label{evolution} \eea Finally, the junction condition on the
dilaton leads to \bea\label{longJu}
\left(H+2\dot{\bar\phi}\right)\left(p+{1\over 2}
T^{\phi}\right)+\mu H=0~. \eea

\section{Brane Cosmology}\setcounter{equation}{0}\label{cosmology}

The cosmological equations seen in the previous section simplify
considerably when the bulk geometry is determined by the AdS
solution found in Sec.~\ref{gensolutions}. Indeed, if we set
$\bar\phi=\bar\phi_0$ and $A=e^{\pm\lambda (z-z_0)}$,
Eqs.~(\ref{longFr})-(\ref{longJu}) reduce respectively to
\bea\label{Fried} H^2&=&{1\over
4}\mu^2-\lambda^2~,\\\non\\\label{consnonstandard} \dot
\mu&=&(\mu+p)H~,\\\non\\\label{juncdilaton} 0&=&\mu+p+{1\over
2}T^{\phi}~.\eea The last equation is always compatible with
Eq.~(\ref{invtfi}) and  with Eq.~(\ref{ident}), hence the set of
transformation rules under T-duality reduce to
\bea\label{redduality} \mu \stackrel{\rm
T}{\longrightarrow}\tilde\mu=-\mu\qquad p\stackrel{\rm
T}{\longrightarrow}\tilde p=p+2\mu,\quad\Leftrightarrow\quad
\omega\stackrel{\rm
T}{\longrightarrow}\tilde\omega=-(2+\omega).\eea The first of
these transformations signals the violation of the strong energy
condition in the duality transition. As we will see below, this
problem can be mitigated by assuming that the total energy density
contains a time-dependent tension which can be positive or
negative, along the lines of \cite{Horowitz}.

From Eq.~(\ref{consnonstandard}), we see that the energy is not
conserved on the brane~\footnote{Note, that our evolution equation
above can also be seen as a special case of the generic set of
energy-non-conserving brane cosmologies studied
in~\cite{Kiritsis:2002zf}.}. However, it is possible to find a new
frame where the energy is conserved. To see this, we define the
conformal induced metric \bea\label{confgamma}
\gamma_{\mu\nu}dx^{\mu}dx^{\nu}=-d\xi^2+E^2\delta_{ij}\,dx^idx^j~,\eea
where $\gamma_{\mu\nu}=e^{\,-4\phi/(D-2)}h_{\mu\nu}~$, and we
require that the fields on the brane couple to $\gamma_{\mu\nu}$
rather than to $h_{\mu\nu}~$. Now, let the Hubble parameter and
the energy momentum tensor, with respect to $\gamma_{\mu\nu}$, be
$~{\cal H}=\partial_{\xi}\ln E~$ and $S_{\mu\nu}$ respectively.
These are related to $H$ and $T_{\mu\nu}$ through the relations
\bea H=-(D-2)e^{\,-2\phi/(D-2)} {\cal H}~, \qquad
T_{\mu\nu}=e^{\,4\phi/(D-2)}S_{\mu\nu}~. \eea It then  follows
that Eq.~(\ref{consnonstandard}) can be written in the standard
form \bea {d\mu\over d\xi} +(D-2)(\mu+p){\cal H}=0~. \eea

We now look at Eqs.~(\ref{Fried})-(\ref{juncdilaton}) for some
relevant values of $\omega=p/\mu$.

\subsection{Pure tension: $\omega=-1$}
In this case $T^{\phi}=0$ and $\mu=\mu_0=\,\,$const, hence
$H^2={\mu_0^2\over 4}-\lambda^2$, which in general corresponds to
an accelerating de Sitter phase. In our case, however, we have
$\mu_0^2=4 \lambda^2$. Indeed, given the definition (\ref{Tphi}),
we see that the condition $T^\phi =0$ implies that ${\cal L} =-
\Lambda\ e^{2\phi(z)}~$, for some constant $\Lambda$. In turn,
this yields $\mu_0 = \Lambda\ e^{2\phi(z(\tau))}~$. Hence, the
vacuum energy density is constant only if the brane is static,
i.e. $\dot z =H=0$. The equality $\mu_0^2=4 \lambda^2$ implies
that \bea \mu_0^2= {16V_0\over
D-1}\,e^{\,\beta\bar\phi_0}~~,\label{selfT} \eea and this shows
that the brane vacuum energy density is arbitrary. In fact, for
any value of the parameter $\beta$, the shifted dilaton ``vacuum''
$\bar\phi_0$ can always be adjusted so that~(\ref{selfT}) is
satisfied, without affecting the Poincar\'{e} invariance of the
solution. Thus, we could consider this as a self tuning mechanism,
pretty much along the same lines of~\cite{KSS}. The main
differences with respect to~\cite{KSS} is that here we work in
string frame and, by including a coupling to the shifted dilaton,
we find a completely smooth, anti-de Sitter bulk. The relation
(\ref{selfT}) appears to leave an ambiguity about the sign of the
energy density. However, by using the $ij$ component of
Eq.~(\ref{sfcurv}) with $T^{\phi}=0$, we find that $2\sigma'=p$.
Hence, for $\sigma=\lambda(z-z_0)$, $p$ is positive and the
tension is negative. On the other hand, $p$ is negative and $\mu$
is positive in the dual case, in agreement with the transformation
laws (\ref{redduality}).

The dual solutions differ not only in the sign of the tension, but
also in the effective Planck mass. By assuming for simplicity that
the brane is located at $z=0$, the latter is given by \bea
M_{PL}^{D-3} \sim \int^0_{-\infty} dz e^{(D-3)\sigma}e^{-2\phi} =
\int^0_{-\infty} dz e^{-2\sigma}. \eea For the solution
$\sigma=\lambda z$, the perturbative expansion (governed by
$e^{\,\phi}$) is under control but the Planck mass diverges, and
gravity is not localised on the brane. The dual solution
$\tilde\sigma=-\lambda z$ yields instead a finite Planck mass, but
the string coupling now is strong. We will comment about these
results in the last section. For the moment, we observe that also
in Jordan frame the solution is static. Indeed, the Friedmann
equation, for pure tension, becomes \bea\label{jordan-f} {\cal
H}^2={\mu_0^2-4\lambda^2\over
4a(D-2)^2}\,e^{\,4\phi(z_b(\tau))/(D-2)}~,\eea where $z_b(\tau)$
denotes the time-dependent location of the brane. This equation
yields again non-static solutions whenever $\mu_0^2
> 4\lambda^2$. However, given that, changing to Jordan frame,
$\mu$ and $p$ are the same, we still have the condition $\dot
z_b=0$, which in turn implies that $\mu_0^2= 4\lambda^2$.

In conclusion, when $\omega=-1~$, only static solutions are
allowed. The vacuum energy on the brane is determined in terms of
the bulk AdS radius through $\lambda$, and it is invariant under
T-duality. Since $\lambda\neq 0$, the brane vacuum energy cannot
vanish, but the solution is always Poincar\'{e} invariant.

\subsection{Perfect fluid: $\omega\neq -1$}
In this case, we have $T^\phi =-2\mu(1+\omega)$ and the evolution
equation yields \bea \mu =\mu_0 \,e^{\,(1+\omega)\sigma}. \eea
This can be plugged into the Friedmann equation (\ref{Fried}),
leading to \bea\label{redFrie} \dot z^{\,2}_b(\tau)={\mu_0^2\over
4\lambda^2}\,e^{\,2(1+\omega)\lambda(z_b(\tau)-z_0)}-1.\eea When
the brane moves towards increasing values of $z$, and for
$\sigma=\lambda(z_{b}(\tau)-z_0)$, we find the solution
\bea\label{zeta} z_b(\tau)-z_0&=&{1\over
2\lambda(1+\omega)}\ln\left\{{4\lambda^2\over
\mu_0^2}\left[1+\tan^2((1+\omega)\lambda(\tau-\tau_0))
\right]\right\},\eea for arbitrary $\tau_0$. It follows that the
scale factor and the energy density are given respectively by \bea
\label{sol:A} A(\tau) &=&\left({2\lambda\over
|\mu_0|}\right)^{1\over 1+\omega}
\biggl[1+\tan^2((1+\omega)\lambda(\tau-\tau_0))\biggr]^{1\over
2(1+\omega)}~,\\\label{sol:mu} \mu(\tau) &=&-2\lambda
\biggl[1+\tan^2((1+\omega)\lambda(\tau-\tau_0))\biggr]^{1\over
2}~. \eea We note that when $\omega>-1$, the scale factor has an
absolute minimum at $\tau=\tau_0$, which corresponds to
$\mu(\tau_0)=-2\lambda$. If we were to make the further assumption
$\mu(\tau_0)\equiv\mu_0$, then $A(\tau_0)=1$ and the bounce would
be located at $z_0$. It is easy to see that this solution
corresponds to a bouncing brane
cosmology~\footnote{See~\cite{Biswas:2005qr} for a recent detailed
study on string inspired bouncing cosmologies.}. Indeed the scale
factor $A(\tau)$ smoothly goes from a contracting phase to an
expanding one, passing through a positive minimum value at
$\tau=\tau_0$. Also, $\dot H>0$ for all $\tau$, hence the Universe
is superinflating at all times. The string coupling is under
control, but the brane has negative tension.

The dual solution can be easily obtained with the substitutions
$\lambda\rightarrow -\lambda$,~ $\omega\rightarrow \tilde\omega$,
and $\mu_0\rightarrow -\tilde\mu_0=-2\lambda$ in the equations
above. The form of the Friedmann equation does not change, and its
solution is \bea\label{dualzeta} z_b(\tau)-\tilde z_0&=&-{1\over
2\lambda(1+\tilde\omega)}\ln\left[1+\tan^2((1+\tilde\omega)\lambda(\tau-\tau_0))
\right]~.\eea However, the dual scale factor is defined by \bea
\tilde A(\tau)\equiv
A(\tau)^{-1}=\,e^{-\lambda(z_b(\tau)-z_0)},\eea which leads to
\bea\label{sol:Atilde} \tilde A(\tau) =
\biggl[1+\tan^2((1+\tilde\omega)\lambda(\tau-\tau_0))\biggr]^{1\over
2(1+\tilde\omega)}~,\eea while the energy density reads \bea
\tilde\mu(\tau) =2\lambda
\biggl[1+\tan^2((1+\tilde\omega)\lambda(\tau-\tau_0))\biggr]^{1\over
2}~. \eea Again, the dual solution still describes a bouncing
Universe, and, provided $\tilde\omega>-1$, the dual scale factor
reaches the minimum $\tilde A(\tau_0)=1$ at the bounce location
$\tilde z_0$. However, the energy density is now positive, but the
string coupling is strong.

As mentioned above, the duality transformations (\ref{redduality})
imply a violation of the strong energy condition. As in the static
case, the solution associated to $\sigma(z_b)$ corresponds to a
negative energy density on the brane. To cope with this problem,
we follow the model proposed in \cite{Horowitz}, and we assume
that \bea\label{phenom} \mu=\rho+F~,\qquad p=\pi-F~, \eea where
$\rho$, $\pi$, and $F$ are treated as functions of $\tau$. By
further imposing that $\tilde\rho=\rho$ and $\tilde\pi=\pi~$, we
separate a T-dual invariant matter contribution from a
time-dependent tension. From Eqs.~(\ref{redduality}) it then
follows that $\tilde F=-F-2\rho~$. If also the matter on the brane
behaves as a perfect fluid, i.e. $\pi=\gamma\rho$, then we have
\bea \rho(\tau)={1+\omega\over 1+\gamma}\,\mu(\tau)~,\eea where
$\mu(\tau)~$ is given by Eq.~(\ref{sol:mu}). Thus, provided
$\omega<-1~$, and $\gamma
>-1~$, we ensure that $\rho$ is always positive. When
$\pi=\rho=0~$, the only source of stress is the tension $F(\phi)$,
hence, under T-duality, we simply have $\tilde F=-F~$, as in the
static case. This shows that even in the simplest case, the
insertion of a brane in the background breaks the physical
equivalence of dual bulks mentioned in Sec.~\ref{gensolutions}.

\subsection{Pre-Big Bang Scenario}

We now come to the implementation of the PBB scenario in our
model. For simplicity, we consider the case when
$-\mu_0=\tilde\mu_0=2\lambda$, and we assume that
$\tilde\omega>-1~$. Using the duality identity $-(\omega+1)=\tilde
\omega +1~$, we compare Eqs.~(\ref{zeta}) and (\ref{dualzeta}),
and we find that $z_0=\tilde z_0$. Then, we construct a global
solution to the Friedmann equation such that \bea z_b(\tau) &=&
z_0+{1\over 2\lambda(\omega+1)}\ln
\left[1+\tan^2\left((1+\omega)\lambda(\tau-\tau_0\right))\right]~,\qquad
\tau < \tau_0~,\\\non\\ z_b(\tau) &=& z_0 -{1\over
2\lambda(\tilde\omega+1)}\ln
\left[1+\tan^2\left((1+\tilde\omega)\lambda(\tau-\tau_0)\right)\right]~,\qquad
\tau > \tau_0~, \eea i.e. we patch at the bounce the solutions
(\ref{zeta}) and (\ref{dualzeta}), respectively. The global scale
factor, for a fixed value of $\tilde\omega~$, becomes \bea
A(\tau)= \left\{
\begin{array}{l}
\left[1+\tan^2\left((1+\tilde\omega)\lambda(\tau-\tau_0)\right)\right]^{-1\over
2(1+\tilde\omega)}~,\qquad \tau <\tau_0~,\\\\
\left[1+\tan^2\left((1+\tilde\omega)\lambda(\tau-\tau_0)\right)\right]^{1\over
2(1+\tilde\omega)}~,\qquad \tau >\tau_0~,
\end{array}\right.\eea and we see that it is always growing, with
a smooth inflection point at $\tau=\tau_0~$. Analogously, the
global Hubble parameter reads \bea H(\tau)= \left\{
\begin{array}{rr}
-2\lambda\tan\left((1+\tilde\omega)\lambda(\tau-\tau_0)\right)~,&\qquad
\tau<\tau_0~,\\\\
2\lambda\tan\left((1+\tilde\omega)\lambda(\tau-\tau_0)\right)~,&\qquad\tau>\tau_0~,
\end{array} \right.\eea
and thus $\dot H$ turns from negative to positive at the bounce,
revealing a superinflationary behaviour for $\tau>\tau_0$.
However, exploiting the results of the static case discussed
above, it is clear that brane gravity is not localised in the
pre-bounce phase whereas it is in the post-bounce one. Moreover,
while in the pre-bounce phase the string coupling is weak, in the
post-bounce it becomes strong, and the model might become
unreliable because of string loop corrections. However, note that
if we interpret $\bar\phi$ as the reduced string coupling constant
\cite{Gasp2}, the latter might be under control as
$\bar\phi=\bar\phi_0$.

In fact, in order to achieve localisation of gravity before the
bounce, one would have to modify the Eq.~(\ref{phenom}), as \bea
\mu(\tau)=-\rho(\tau)+F(\tau)~,\qquad
p(\tau)=-\pi(\tau)-F(\tau)~,\eea so that $\tilde\omega<-1$. Then,
the global scale factor and Hubble parameter can be obtained by
the expressions above by simply replacing $\tilde\omega$ with
$\omega~$.

In both cases, this construction violates the positivity of the
total energy density on the brane. Note that this violation is not
localised around $z_0$, but it holds for all times before (or
after) the dual transition, in opposition to the PBB scenario,
where the violation (if any) is always localised.

\section{Discussion}

We have considered a setup where a codimension-one brane is
embedded in a static bulk, and we assumed T-duality symmetry along
the time direction, as well as the space directions parallel to
the brane. The time-like T-duality is shown to be consistent as
the bulk geometry is anti-de Sitter.

When the brane is static, there are smooth solutions, which show
Poincar\'e invariance on the brane, regardless of the value of its
tension. This is possible thanks to a self-tuning property of the
bulk shifted-dilaton vacuum energy. We also found brane-moving
solutions, which lead to both bouncing and PBB-like cosmology. In
the latter case, we found that gravity can be localised on the
brane via T-duality, however the model incurs a violation of the
strong energy condition.

In all these cases, regular solutions for which the string
coupling is under control do not lead to localisation of gravity
on the brane. This is not {\it a priori} bad news. Indeed, gravity
could be {\it induced} on the brane via quantum effects, as
discovered in~\cite{Dvali:2000hr}. Note, in fact, that including
an Einstein-Hilbert term on the brane does not require (at least
in the static case) a modification of the AdS${}_5$ background
studied here, the latter being 4D flat. Moreover, the van
Dam-Veltman-Zakharov discontinuity of the graviton propagator,
present in the 5D flat case~\cite{Dvali:2000hr}, might be absent
here, where the background is AdS~\cite{Porrati:2000cp}. It also
appears that previously found no-go
theorems~\cite{Csaki:2000wz,Cline:2001yt} for self-tuning RS
backgrounds, do not apply here as the background we found is
completely smooth. However, similar setups studied in the past, in
Einstein frame, seem to require conformality of the localised
matter on the brane~\cite{Kakushadze:2000ix}, and it is not a
priori obvious whether such consistency conditions are present in
our setup as well. Another way of inducing 4D gravity on the brane
is via bulk higher-curvature terms~\cite{Corradini:2001qv};
however, this might spoil the self-tuning feature in such a
codimension-one model~\cite{Low,Binetruy:2002ck}.

Besides these issues, the model studied in this Letter can be
extended in other ways. For instance, one can consider the
embedding of a second brane, which might regularise the effective
Planck mass. Also, a second brane could cure the singularities,
mentioned in Sec.~(\ref{gensolutions}), that appear in the bulk
when the shifted dilaton is not constant. Finally, one might
consider including in the bulk the NS anti-symmetric tensor field,
along the lines of some PBB models.

Generally speaking, time-like T-duality is a symmetry not yet
fully explored in the context of string cosmology, and it might be
worth investigating. As an example, the inclusion of time-like
duality in the String Gas Cosmology of Brandenberger and Vafa
\cite{Brand}, might lead to interesting insights.

\acknowledgments{The work of OC has been partly supported by the
EC commission
  via the FP5 grant HPRN-CT-2002-00325. OC is grateful to the University of
  Padova for financial support via the PRIN-2003 project. We wish to thank
  D.~Wands and M.~Gasperini for useful suggestions, and P.~Watts for reading
  the manuscript.}

\appendix

\section{Extrinsic curvature and its dual}
\label{appendix:extrinsic}

In this appendix we study how the extrinsic curvature transforms
under T-duality. We consider a brane with (D-1)-dimensional
induced metric \bea
h_{\mu\nu}dx^{\mu}dx^{\nu}=-d\tau^2+e^{2\sigma(z)}\delta_{ij}dx^idx^j~,\eea
embedded in a D-dimensional bulk space-time with metric \bea
g_{AB}\,dx^Adx^B=e^{2\sigma(z)}\eta_{\mu\nu}dx^{\mu}dx^{\nu}+dz^2~,\eea
where $\eta_{\mu\nu}$ denotes a (D-1)-dimensional Minkowski
metric. We follow the conventions of \cite{Brax}, and we assume
that the brane moves in the bulk with velocity $v^{A}=(\dot t,
0,0,\ldots,0,\dot z_b)$, where $z_b(\tau)~$ is the brane location
in the bulk, and the dot stands for differentiation with respect
to $\tau$. Our choice for the induced metric implies that $\dot
t=\pm e^{-\sigma}\sqrt{1+\dot z_b^2}~$. The vector $n^A$ normal to
the brane is defined such that \bea g_{AB}\,n^An^B=1~,\qquad
n_Av^A=0~.\eea These conditions yield the components of $n^A$,
namely \bea n_{A}=\left(\pm e^{\sigma}\dot
z_b,0,0,\ldots,0,\mp\sqrt{1+\dot z_b^2}\right)~.\eea If we choose
$(+,-)$, we ensure that the normal vector points into the
bulk~\footnote{Here we also choose $\dot t>0$, and this leads to
the condition sign$(n^t)=$ sign$(n^z)$. If instead one takes $\dot
t<0$, then sign$(n^t)=-$sign$(n^z)$.}. The extrinsic curvature is
defined as \bea K_{\mu\nu}={\partial X^A\over \partial
x^{\mu}}{\partial X^B\over\partial x^{\nu}}\nabla_An_B~~,\eea
where $X^A$ ($x^{\mu}$) are the bulk (brane) coordinates. In our
case, its components are given by \bea K_{ij}=-\sigma'\sqrt{1+\dot
z_b^2}\, h_{ij}~,\qquad K_{\tau\tau}=e^{-\sigma}{d\over dz_b}\,
e^{\sigma}\sqrt{1+\dot z_b^2}~,\eea hence the trace reads \bea
K_{\mu\nu}h^{\mu\nu}= K= -(D-1)\sigma'\sqrt{1+\dot z_b^2}-{d\over
dz_b}\sqrt{1+\dot z_b^2}~. \eea

The T-dual bulk and induced metrics considered in this work are
obtained by replacing $\sigma$ with $-\sigma$. The velocity vector
can still be defined as $v^{A}=(\dot t, 0,0,\ldots,0,\dot z_b)$,
while the normalisation condition on $\dot t$ now becomes $\dot
t=\pm e^{\sigma}\sqrt{1+\dot z_b^2}~$. By repeating the same
computations as above (with the same choice for the signs) it is
easy to show that the dual normal vector becomes \bea \tilde
n_{A}=(e^{-\sigma}\dot z_b,0,0,\ldots,0,-\sqrt{1+\dot
z_b^2})~,\eea while the components of the extrinsic curvature read
\bea \tilde K_{ij}=\sigma'\sqrt{1+\dot z_b^2}\tilde h_{ij}~,\qquad
\tilde K_{\tau\tau}=e^{\sigma}{d\over dz_b}e^{-\sigma}\sqrt{1+\dot
z_b^2}~.\eea Thus, the relevant T-duality transformations read
\bea n_t&\stackrel{\rm T}{\longrightarrow}&\tilde
n_t=e^{-2\sigma}n_t~,
\qquad n_z\stackrel{\rm T}{\longrightarrow}\tilde n_z=n_z~,\\\non\\
K_{ij}&\stackrel{\rm T}{\longrightarrow}&\tilde
K_{ij}=-e^{-4\sigma}K_{ij}~,
\\\non\\
\tilde K_{\tau\tau}&\stackrel{\rm T}{\longrightarrow}&\tilde
K_{\tau\tau}=
K_{\tau\tau}-2\sigma'\sqrt{1+\dot z_b^2}~,\\\non\\
K&\stackrel{\rm T}{\longrightarrow}&\tilde
K=K+2(D-1)\sigma'\sqrt{1+\dot z_b^2}~.\eea

\end{document}